\documentclass[conference, 10pt]{IEEEtran}
\IEEEoverridecommandlockouts
\usepackage{cite}
\usepackage{amsmath,amssymb,amsfonts}
\usepackage{algpseudocode}
\usepackage{graphicx}
\usepackage{textcomp}
\usepackage{xcolor}
\usepackage{braket}
\usepackage{url}
\usepackage{booktabs} 
\usepackage{multirow} 

\newcommand{\figureautorefname}{Figure~\negthinspace}

\newcommand{\tableautorefname}{Table~\negthinspace}

\def\BibTeX{{\rm B\kern-.05em{\sc i\kern-.025em b}\kern-.08em
    T\kern-.1667em\lower.7ex\hbox{E}\kern-.125emX}}
\begin{document}

\title{Federated Quantum-Train Long Short-Term Memory for Gravitational Wave Signal \thanks{The views expressed in this article are those of the authors and do not represent the views of Wells Fargo. This article is for informational purposes only. Nothing contained in this article should be construed as investment advice. Wells Fargo makes no express or implied warranties and expressly disclaims all legal, tax, and accounting implications related to this article.}
}


\author{
\IEEEauthorblockN{
     Chen-Yu Liu \IEEEauthorrefmark{1}\IEEEauthorrefmark{6}\IEEEauthorrefmark{9}, 
    Samuel Yen-Chi Chen\IEEEauthorrefmark{3},
    Kuan-Cheng Chen\IEEEauthorrefmark{4}\IEEEauthorrefmark{5}\IEEEauthorrefmark{8},
    Wei-Jia Huang\IEEEauthorrefmark{9}, 
    Yen-Jui Chang \IEEEauthorrefmark{10}\IEEEauthorrefmark{11}
}

\IEEEauthorblockA{\IEEEauthorrefmark{1}Graduate Institute of Applied Physics, National Taiwan University, Taipei, Taiwan}
\IEEEauthorblockA{\IEEEauthorrefmark{9}Hon Hai (Foxconn) Research Institute, Taipei, Taiwan}
\IEEEauthorblockA{\IEEEauthorrefmark{3}Wells Fargo, New York, NY, USA}
\IEEEauthorblockA{\IEEEauthorrefmark{4}Department of Electrical and Electronic Engineering, Imperial College London, London, UK}
\IEEEauthorblockA{\IEEEauthorrefmark{5}Centre for Quantum Engineering, Science and Technology (QuEST), Imperial College London, London, UK}

\IEEEauthorblockA{\IEEEauthorrefmark{10}Quantum Information Center, Chung Yuan Christian University, Taoyuan City, Taiwan}
\IEEEauthorblockA{\IEEEauthorrefmark{11}Master Program in Intelligent Computing and Big Data, Chung Yuan Christian University, Taoyuan City, Taiwan}
\IEEEauthorblockA{Email:\IEEEauthorrefmark{6} d10245003@g.ntu.edu.tw, \IEEEauthorrefmark{8}kuan-cheng.chen17@imperial.ac.uk}

}

\maketitle
\begin{abstract}
We present Federated QT-LSTM, a novel framework that combines the Quantum-Train (QT) methodology with Long Short-Term Memory (LSTM) networks in a federated learning setup. By leveraging quantum neural networks (QNNs) to generate classical LSTM model parameters during training, the framework effectively addresses challenges in model compression, scalability, and computational efficiency. Importantly, Federated QT-LSTM eliminates the reliance on quantum devices during inference, making it practical for real-world applications. Experiments on simulated gravitational wave (GW) signal datasets demonstrate the framework’s superior performance compared to baseline models, including LSTM and QLSTM, achieving lower training and testing losses while significantly reducing the number of trainable parameters. The results also reveal that deeper QT layers enhance model expressiveness for complex tasks, highlighting the adaptability of the framework. Federated QT-LSTM provides a scalable and efficient solution for privacy-preserving distributed learning, showcasing the potential of quantum-inspired techniques in advancing time-series prediction and signal reconstruction tasks.
\end{abstract}

\begin{IEEEkeywords}
Quantum Machine Learning, Quantum Neural Networks, Model Compression
\end{IEEEkeywords}

\section{Introduction}

Quantum Computing (QC) and Quantum Machine Learning (QML) are rapidly evolving fields poised to redefine computational methodologies across a wide range of applications \cite{biamonte2017quantum}. By harnessing intrinsic quantum properties such as superposition and entanglement, QML algorithms facilitate concurrent information processing across numerous states, delivering a level of parallelism that remains unattainable by classical computing architectures \cite{lau2017quantum}.
%
%
This capability has paved the way for innovative applications, including classification tasks \cite{mitarai2018quantum,chen2021end,chen2022quantumCNN,qmlapp2,chen2024compressedmediq}, reinforcement learning \cite{chen2020variational,chen2022variational,yun2023quantum,chen2024efficient}, time-series forecasting \cite{chen2022quantumLSTM,chen2022reservoir,lin2024quantum}, and, notably, the modeling and prediction of gravitational waves—an essential challenge in high-energy physics and cosmology \cite{chen2024qcq}.

%
Although effective, QML algorithms depend on high-quality training datasets, which are often proprietary or reside on edge devices, making the transfer of these datasets to centralized quantum cloud services undesirable or impractical. Frameworks such as \emph{Quantum Federated Learning} (QFL) and \emph{Distributed Quantum Computing} enable the utilization of quantum computational power while preserving data locality on edge devices \cite{chen2021federated,chehimi2022quantum,kwak2023quantum,chehimi2023foundations,rofougaran2024federated,Chehimi2024FedQLSTM,chu2023cryptoqfl,qu2024quantum}.
%
Yet, conventional QML approaches face significant challenges when encoding large datasets into quantum systems. Methods like gate-angle encoding and amplitude encoding \cite{huang2021power}, though theoretically promising, are hindered by the limited qubit count and short coherence times of current quantum hardware. Additionally, relying on cloud-based quantum services during the inference phase introduces latency and inefficiencies, posing critical limitations for mission-critical autonomous systems.
The Quantum-Train (QT) framework introduces a hybrid quantum-classical architecture to address data encoding challenges \cite{liu2024training, liu2024quantum, liu2024qtrl, lin2024quantum, liu2024federated, liu2024quantum2, lin2024quantum2, liu2024quantum3}. 
The QT approach uses a quantum neural network (QNN) solely to generate weights for a classical model, bypassing the data encoding bottleneck. Post-training, the model functions entirely classically, eliminating reliance on quantum hardware during inference and ensuring efficiency and practicality for near-term QML applications.
%

This paper presents the Federated Quantum-Train Long Short-Term Memory (Fed-QT-LSTM) as an innovative solution to enhance QML efficiency in edge computing while enabling sequential data processing with memory-capable models. Numerical simulations demonstrate that Fed-QT-LSTM excels in gravitational wave signal data, surpassing QLSTM of similar model size and achieving performance comparable to larger classical LSTM models.

\begin{figure*}[htbp]
\vskip -0.15in
\begin{center}
\includegraphics[width=2\columnwidth]{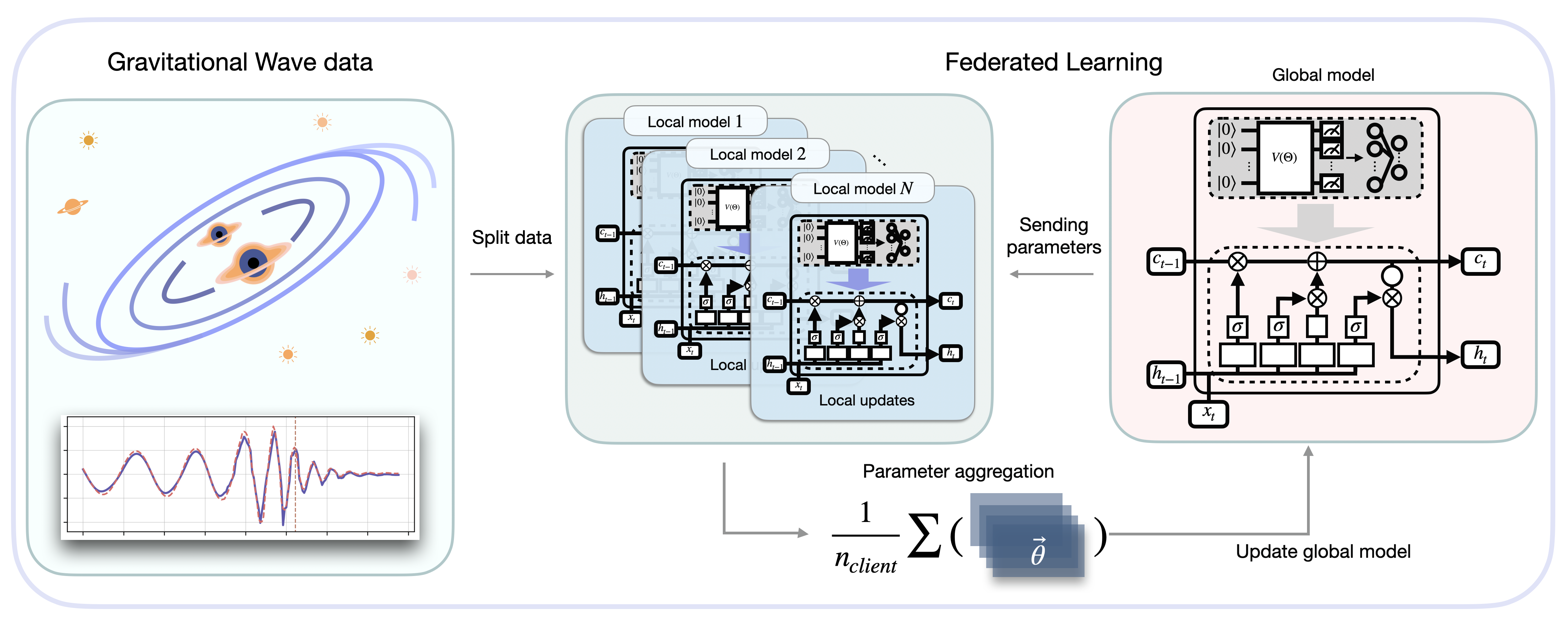}\vskip -0.1in
\caption{{\bfseries Scheme: Fed-QT-LSTM framework for Gravitational Wave Signal.
}}
\label{fig:scheme}
\end{center}
\vskip -0.25in
\end{figure*}

\section{Related Works}
\label{sec:related_works}
%
%
\emph{Long Short-Term Memory} (LSTM) networks \cite{hochreiter1997long} represent a foundational recurrent neural network (RNN) architecture widely utilized for sequential modeling tasks, including time-series prediction and machine translation \cite{sutskever2014sequence}. In recent developments, Quantum LSTM (QLSTM) models have been proposed to enhance LSTM capabilities by leveraging QNNs \cite{chen2022quantumLSTM}. QLSTM has demonstrated success across various applications, such as time-series forecasting \cite{chen2022quantumLSTM,chen2022reservoir} and natural language processing \cite{li2023pqlm,stein2023applying,di2022dawn}. Unlike prior QLSTM \cite{chen2022quantumLSTM} and classical LSTM approaches \cite{hochreiter1997long}, this paper introduces a novel application of QT for generating LSTM model weights, thereby eliminating the need for encoding classical input data into quantum systems. Furthermore, this work distinguishes itself from existing Quantum-Train LSTM (QT-LSTM) frameworks \cite{lin2024quantum} by addressing federated or distributed learning scenarios. Specifically, this paper presents an innovative framework that integrates quantum federated learning, quantum sequence learning, and quantum-enhanced neural network weight generation.
\section{Quantum Neural Networks}
\begin{figure}[htbp]
\vskip -0.1in
\begin{center}
\includegraphics[width=0.85\columnwidth]{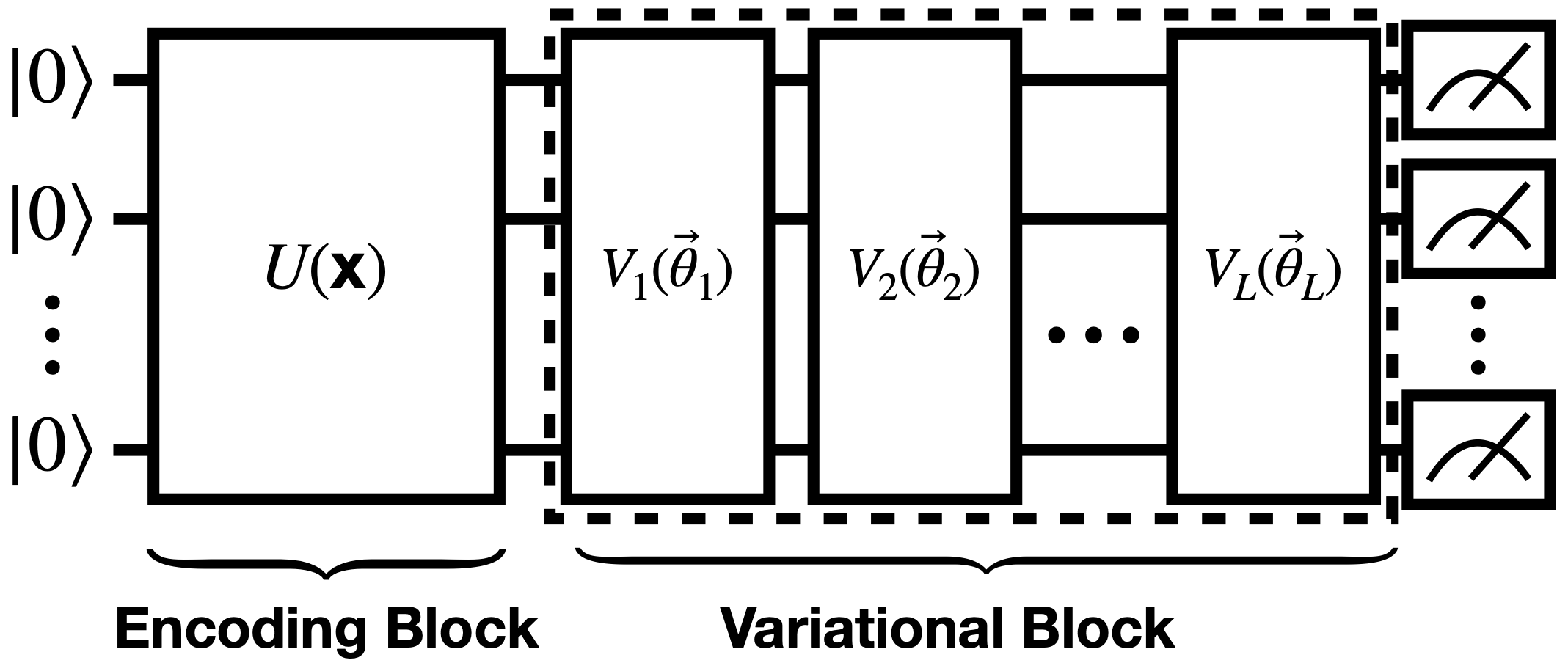}\vskip -0.1in
\caption{{\bfseries Generic Structure of a Variational Quantum Circuit (VQC).}}
\label{fig:generic_VQC}
\end{center}
\vskip -0.1in
\end{figure}

The design of QNNs within the contemporary hybrid quantum-classical computing paradigm is primarily driven by \emph{variational quantum circuits} (VQCs), also referred to as \emph{parameterized quantum circuits} (PQCs). These circuits form a distinct class of quantum architectures featuring trainable parameters. Recent studies have demonstrated that VQCs or PQCs can achieve specific forms of quantum advantage \cite{abbas2021power,yu2024shedding,du2020expressive}, underscoring their potential to significantly enhance QML applications.
%
%
There are three core components in a VQC or PQC: an \emph{encoding circuit}, a \emph{variational circuit}, and the concluding \emph{measurements}.
%
%
%
As illustrated in \figureautorefname{\ref{fig:generic_VQC}}, the encoding circuit $U(\mathbf{x})$ maps the initial quantum state $\ket{0}^{\otimes n}$ to the state $\ket{\Psi} = U(\mathbf{x})\ket{0}^{\otimes n}$. Here, $n$ denotes the number of qubits, and $U(\mathbf{x})$ is a unitary operator parameterized by the input vector $\mathbf{x}$.

\begin{figure}[htbp]
\vskip -0.1in
\begin{center}
\includegraphics[width=0.85\columnwidth]{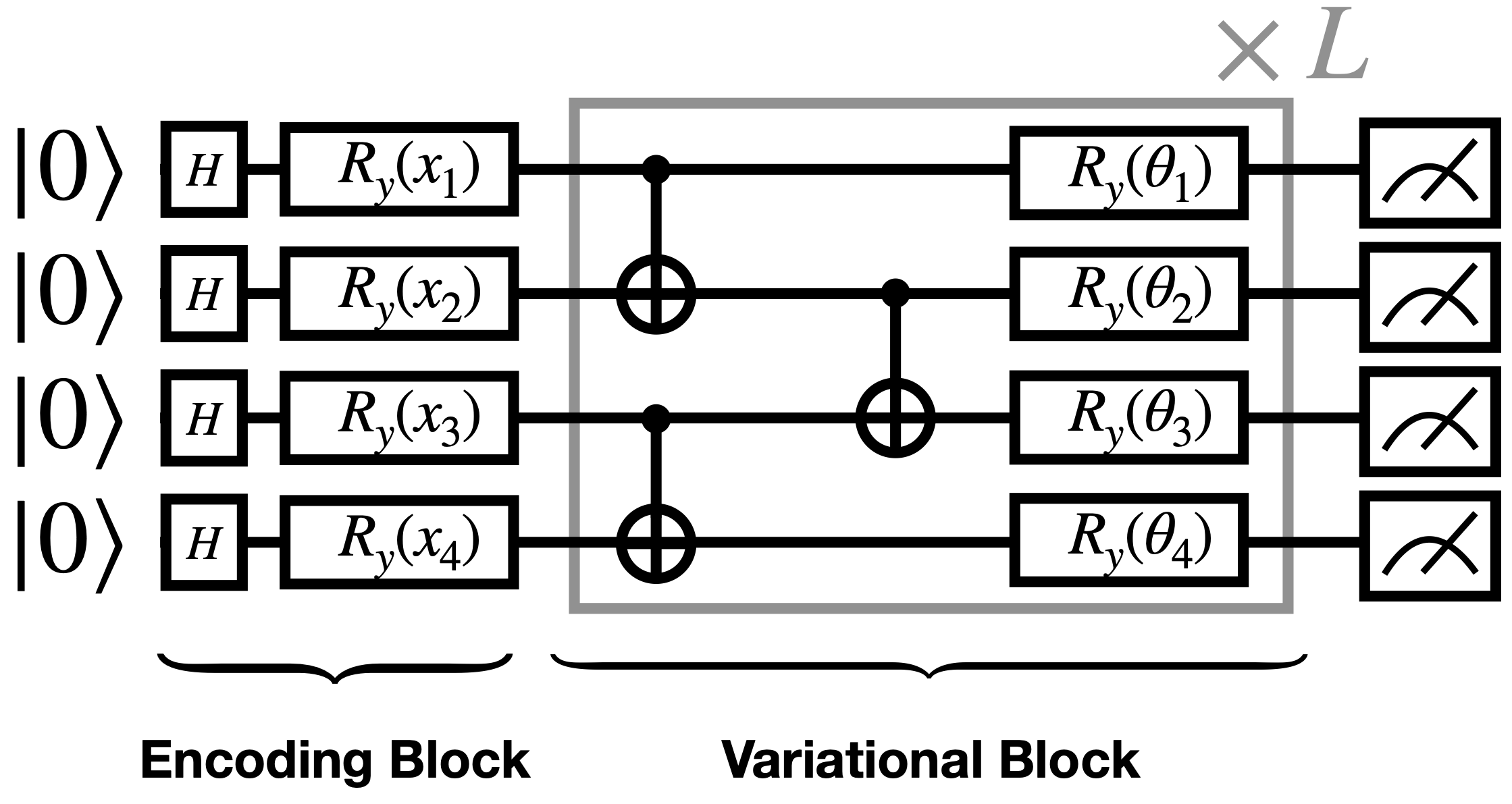}\vskip -0.1in
\caption{{\bfseries VQC used in this paper.}}
\label{fig:detailed_VQC}
\end{center}
\vskip -0.2in
\end{figure}

%
%
The VQC for the \emph{QLSTM} utilized in this study is illustrated in \figureautorefname{\ref{fig:detailed_VQC}}. For illustration purposes, a 4-qubit system is used.
%
%
The encoding circuit block $U(\mathbf{x})$ begins by applying Hadamard gates to all qubits, creating an unbiased superposition state $H\ket{0} \otimes \cdots \otimes H\ket{0} = \sum_{(q_1, q_2, \dots, q_n) \in \{0,1\}^n} \frac{1}{\sqrt{2^n}} \ket{q_1} \otimes \ket{q_2} \otimes \cdots \otimes \ket{q_n}$. Following this, $R_y$ gates are applied, where the rotation angles are defined by the input data $x_1, \dots, x_n$. This results in the encoding operation $U(\mathbf{x}) = R_y(x_1)H \otimes \cdots \otimes R_y(x_n)H$.

%
%
The encoded state is subsequently processed by the variational block (indicated by the dashed box in \figureautorefname{\ref{fig:generic_VQC}}), which consists of multiple layers of trainable quantum circuits $V_{j}(\vec{\theta_{j}})$. The specific $V_{j}$ circuit block employed in this study is shown in the boxed region of \figureautorefname{\ref{fig:detailed_VQC}}. To enhance the expressiveness and increase the number of trainable parameters, these layers can be repeated $L$ times. Each $V_{j}$ block includes CNOT gates for entangling quantum information and parameterized $R_{y}$ gates.
%
%
The trainable component of the circuit is denoted as $W(\Theta)$, defined as $W(\Theta) = V_{L}(\Vec{\theta_{L}})V_{L-1}(\Vec{\theta_{L-1}}) \cdots V_{1}(\Vec{\theta_{1}})$, where $L$ is the number of layers and $\Theta$ represents the set of trainable parameters $ \Vec{\theta_{1}}, \cdots, \Vec{\theta_{L}}$. Each $\Vec{\theta_{k}}$ corresponds to the parameter subset $(\theta_{1}, \cdots, \theta_{n})$ for the respective layer.

%
%
The VQC for the \emph{Quantum-Train} module, designed to generate classical neural network weights for the LSTM, follows the same architecture, incorporating Hadamard gates for initialization and the variational blocks described earlier. The key difference in the \emph{Quantum-Train} VQC module is the exclusion of $R_y$ gates used for data encoding.
%
In the \emph{QLSTM} VQC, Pauli-$Z$ expectation values are retrieved during the measurement process. The output can be expressed as $\overrightarrow{f(\mathbf{x} ; \Theta)}=\left(\left\langle\hat{Z}_1\right\rangle, \cdots,\left\langle\hat{Z}_n\right\rangle\right)$, where $\left\langle\hat{Z}_{k}\right\rangle =\left\langle 0\left|U^{\dagger}(\mathbf{x})W^{\dagger}(\Theta) \hat{Z}_{k} W(\Theta)U(\mathbf{x})\right| 0\right\rangle$. 
%
Conversely, in the \emph{Quantum-Train} VQC module, the probabilities of all $2^{n}$ computational basis states ($\ket{00 \cdots 0}, \cdots, \ket{11 \cdots 1}$) are extracted and utilized to generate the neural network weights.
%
%
%
%
\section{Federated Learning}
\label{sec:fl}
%
\emph{Federated Learning} (FL) \cite{mcmahan2017communication} has recently emerged as a promising framework to address challenges in privacy-preserving deep learning and facilitate large-scale learning across multiple participants. The fundamental components of an FL system include a \emph{central node} and multiple \emph{client nodes}. The central node maintains the global model and collects trained parameters from the client devices. It performs an \emph{aggregation} process to update the global model and subsequently distributes the updated model to all client nodes. Each client node trains the received model locally using its respective subset of data, which typically represents only a small fraction of the overall dataset.
\section{Long Short-term Memory}
\label{sec:lstm}
%
QLSTM is a quantum adaptation of the traditional LSTM, where the classical neural networks in the LSTM are replaced with QNNs or VQCs \cite{chen2022quantumLSTM}. The core operations, including the gating mechanisms within a QLSTM cell, remain identical to those of its classical counterpart.
\begin{figure}[htbp]
\begin{center}
\includegraphics[width=0.8\columnwidth]{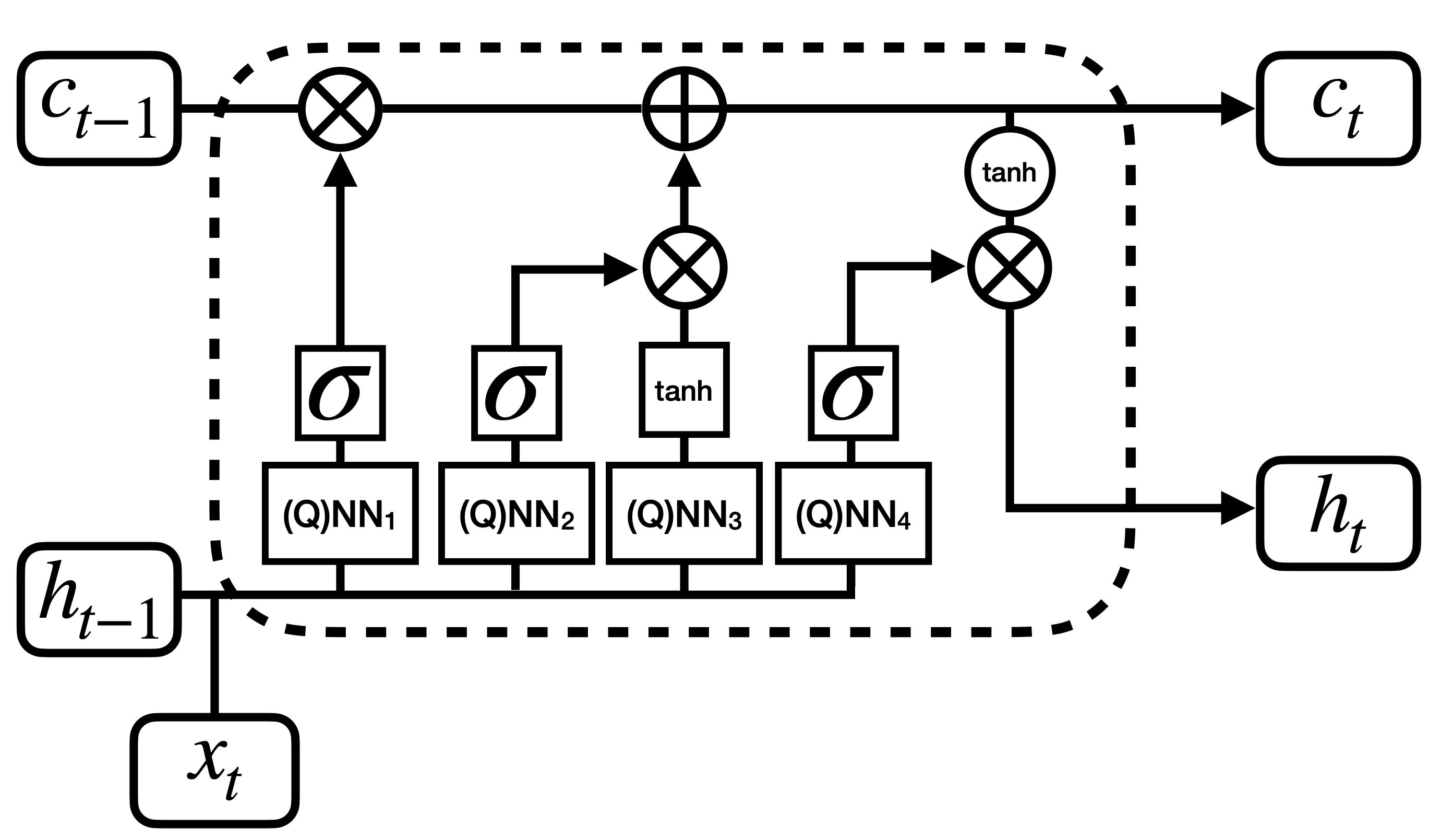}\vskip -0.1in
\caption{{\bfseries (Quantum) Long Short-term Memory.}}
\label{fig:LSTM_and_QLSTM}
\end{center}
\vskip -0.2in
\end{figure}
A formal mathematical formulation of a (Q)LSTM cell is given by,
\begin{subequations}
\allowdisplaybreaks
    \begin{align}
    f_{t} &= \sigma\left(\text{(Q)NN}_{1}(v_t)\right) \label{eqn:qlstm-f}\\
    i_{t} &= \sigma\left(\text{(Q)NN}_{2}(v_t)\right) \label{eqn:qlstm-i}\\ 
    \tilde{C}_{t} &= \tanh \left(\text{(Q)NN}_{3}(v_t)\right) \label{eqn:qlstm-bigC}\\
    c_{t} &= f_{t} * c_{t-1} + i_{t} * \tilde{C}_{t} \label{eqn:qlstm-c}\\
    o_{t} &= \sigma\left(\text{(Q)NN}_{4}(v_t)\right) \label{eqn:qlstm-o}\\ 
    h_{t} &= o_{t} * \tanh \left(c_{t}\right) \label{eqn:qlstm-h}
    \end{align}
    \label{eqn:qlstm}
\end{subequations}
where $v_t=\left[h_{t-1}, x_{t}\right]$ represents the concatenation of input $x_{t}$ at time-step $t$ and the hidden state $h_{t-1}$ from the previous time-step $t-1$.
\section{Quantum-Train}
For a classical model with $p$ parameters, denoted as $\kappa \in \mathbb{R}^{p}$, the QT framework utilizes a QNN with $n_{qt} = \lceil \log_{2} p \rceil$ qubits. This QNN creates a Hilbert space of dimension $2^{\lceil \log_{2} p \rceil}$, which is sufficient to generate $2^{n_{qt}}$ unique measurement probabilities. These probabilities are represented as $|\langle \phi_i | \psi(\gamma) \rangle|^2 \in [0,1]$, where $|\phi_i \rangle$ denotes the basis states for $i \in \{1,2, \ldots, 2^{n_{qt}} \}$. To bridge the quantum probabilities and the classical model parameters, a mapping function $\mathcal{M}_{\beta}$, parameterized by $\beta$, is introduced. This mapping function, implemented as a classical multi-layer perceptron (MLP), maps the basis states $|\phi_i \rangle$ (represented as binary strings) and the corresponding measurement probabilities $|\langle \phi_i | \psi(\gamma) \rangle|^2$ to the target parameter values $\kappa_i$:
\[
\mathcal{M}_{\beta}(|\phi_i \rangle, |\langle \phi_i | \psi(\gamma) \rangle|^2) = \kappa_i, \quad \forall i \in \{1,2, \ldots, p\}.
\]
Consequently, the training process involves optimizing both the QNN parameters $\gamma$ and the mapping model parameters $\beta$. Given that both $|\psi(\gamma)\rangle$ and $ \mathcal{M}_{\beta}$ require only $O(\mathrm{poly}(n_{qt}))$ parameters \cite{cerezo2021variational}, QT achieves a significant reduction in the number of parameters from $p$ to $O(\mathrm{polylog}(p))$.

\subsection{QT-LSTM}
\begin{figure}[htbp]
\vskip -0.2in
\begin{center}
\includegraphics[width=0.8\columnwidth]{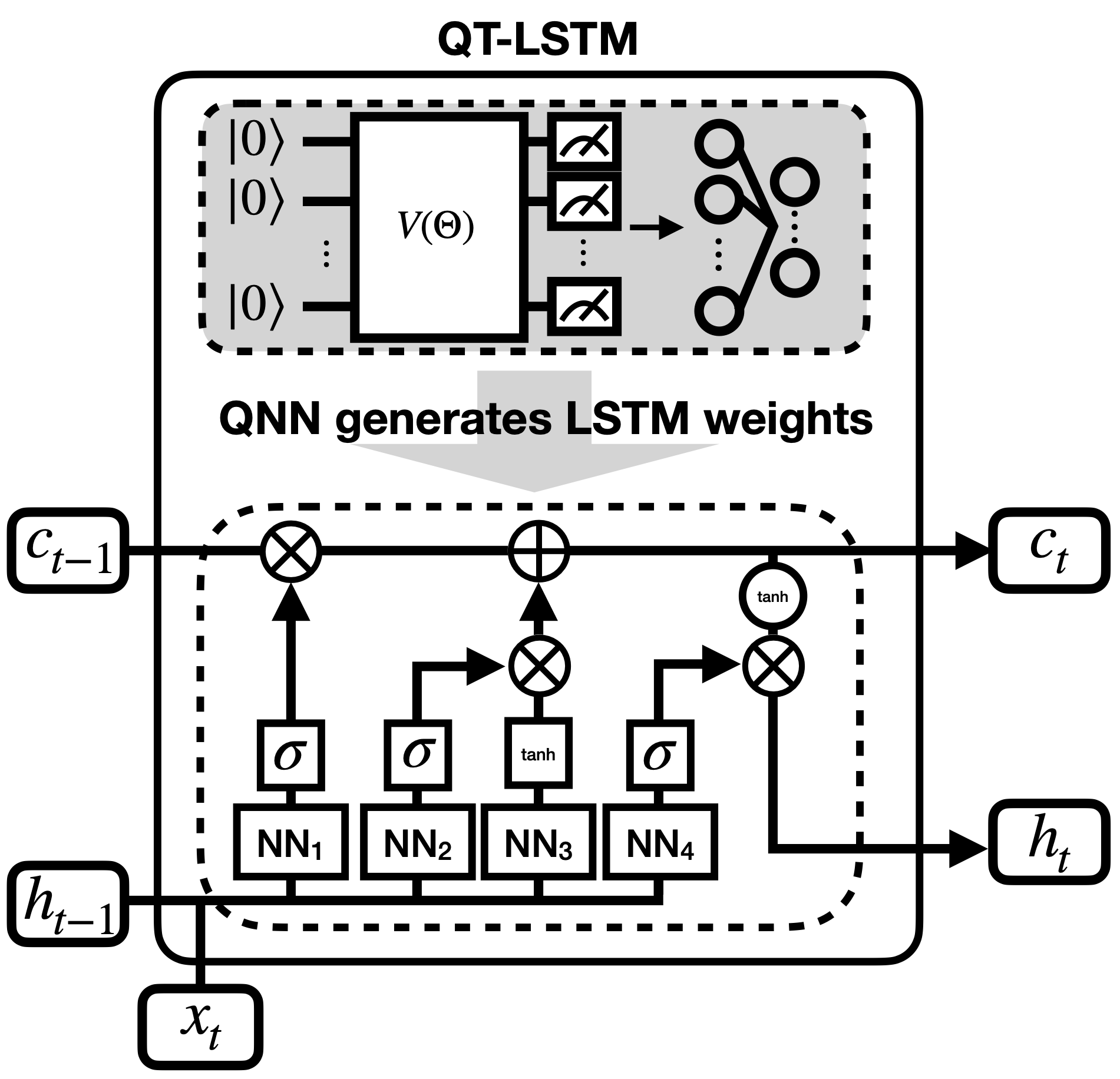}\vskip -0.1in
\caption{{\bfseries Quantum-Train LSTM (QT-LSTM).}}
\label{fig:QT_LSTM}
\end{center}
\vskip -0.2in
\end{figure}

%
%
The original LSTM \cite{hochreiter1997long} architecture relies on the direct training of NN weights within the LSTM cell. In contrast, the QLSTM \cite{chen2022quantumLSTM} model leverages gradient-based optimization of QNN parameters. While QLSTM demonstrates superior performance compared to classical LSTM in terms of training efficiency, a significant challenge is its reliance on quantum devices during the inference stage. To address this limitation, QT-LSTM has been proposed as a framework that combines the strengths of both classical and quantum computing paradigms, as described in \cite{lin2024quantum}. The core concept involves utilizing a QNN to generate the NN weights for the LSTM during the training phase. After training, only the generated NN weights are retained for inference, thereby eliminating the need for continuous access to quantum devices while still benefiting from the enhanced training efficiency provided by the QNN. We illustrate the concept of QT-LSTM in \figureautorefname{\ref{fig:QT_LSTM}}.
%
\section{Numerical Results and Discussion}
\label{sec:nrd}


To evaluate the effectiveness of the Fed-QT-LSTM and compare its performance with Federated LSTM and QLSTM, the Gravitational Waves (GWs) dataset is utilized. GWs are ripples in the fabric of spacetime caused by accelerating massive objects, such as merging black holes or neutron stars\cite{abbott2016observation}. The GW signals used in this study are simulated waveforms from binary black hole mergers, generated using inspiral-merger-ringdown models consistent with General Relativity, as detailed on the Kaggle page for the Riroriro GW simulation package\cite{van2021riroriro}. The goal is to reconstruct the GW signals from the data using the Fed-QT-LSTM. Achieving success in this task highlights the algorithm’s potential to contribute significantly to GW astronomy by improving signal detection and parameter estimation capabilities. 


Following the foundational concepts of FL \cite{mcmahan2017communication}, QFL \cite{chen2021federated}, and Federated QT \cite{liu2024federated}, we extend the Federated QT framework from a naive CNN model to an LSTM model. In this approach, each quantum client node employs a QNN $|\psi (\theta) \rangle$ along with a mapping model $\Tilde{G}_{\beta}$ to generate the local target LSTM model parameters. During each training round, quantum client nodes update their QNN parameters and associated mapping model parameters based on their local datasets. These updated parameters are then transmitted to a central node, where they are aggregated to update the global model through classical channel, as illustrated in  \figureautorefname{\ref{fig:scheme}}. This iterative process ensures that the global model benefits from the quantum training performed at each client node, resulting in enhanced performance and efficiency. By integrating the QT-LSTM model into FL, we leverage quantum computing to reduce the number of training parameters and improve the scalability of distributed learning systems. Importantly, unlike traditional QFL, Fed-QT-LSTM does not require a quantum computer during the inference stage.

As shown in \tableautorefname{\ref{tab:parameters}}, for the comparison baselines, the Federated LSTM model utilizes 1,781 parameters, the Federated QLSTM utilizes 205 parameters, and the Fed-QT-LSTM uses 123 parameters to generate the classical LSTM model with 1,781 parameters during training, in the case of QT layers $ L = 10 $. In all federated learning schemes, 100 training rounds are performed with 4 local clients, each training for 1 local epoch. The global model is aggregated using the averaging method. Notably, the QT-LSTM employs $ n_{qt} = \lceil \log_2 1781 \rceil = 11 $ qubits to compress the information of the classical LSTM model into the quantum-train framework.

\begin{table}[]
\caption{Number of trainable parameters in LSTM, QLSTM, and QT-LSTM models in federated learning scheme.}
\label{tab:parameters}
\resizebox{\columnwidth}{!}{%
\begin{tabular}{|cc|cc|cc|}
\hline
\multicolumn{2}{|c|}{LSTM}                & \multicolumn{2}{c|}{QLSTM \cite{chen2022reservoir}} & \multicolumn{2}{c|}{QT-LSTM ($L=10$)}    \\ \hline
\multicolumn{1}{|c|}{Classical} & Quantum & \multicolumn{1}{c|}{Classical}               & Quantum               & \multicolumn{1}{c|}{Classical} & Quantum \\ \hline
\multicolumn{1}{|c|}{1781}      & --      & \multicolumn{1}{c|}{5}                       & 200                   & \multicolumn{1}{c|}{13}        & 110     \\ \hline
\end{tabular}%
}
\end{table}

\begin{figure}[htbp]
\vskip -0.15in
\begin{center}
\includegraphics[width=0.95\columnwidth]{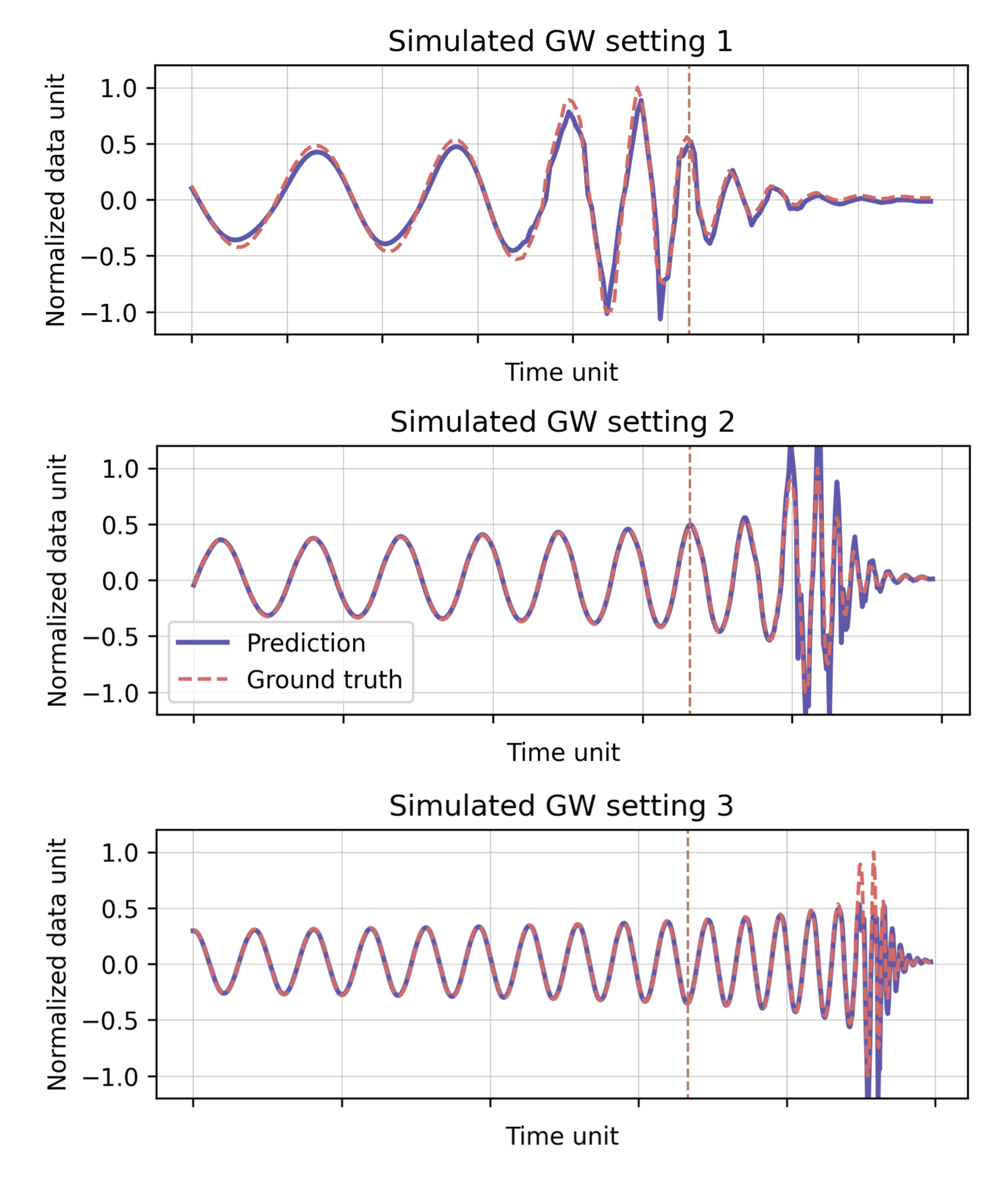}\vskip -0.1in
\caption{{\bfseries Results: Different settings of simulated GW, with prediction of federated QT-LSTM.}}
\label{fig:simu_res_1}
\end{center}
\vskip -0.25in
\end{figure}

\begin{figure*}[htbp]
\vskip -0.1in
\begin{center}
\includegraphics[width=1.8\columnwidth]{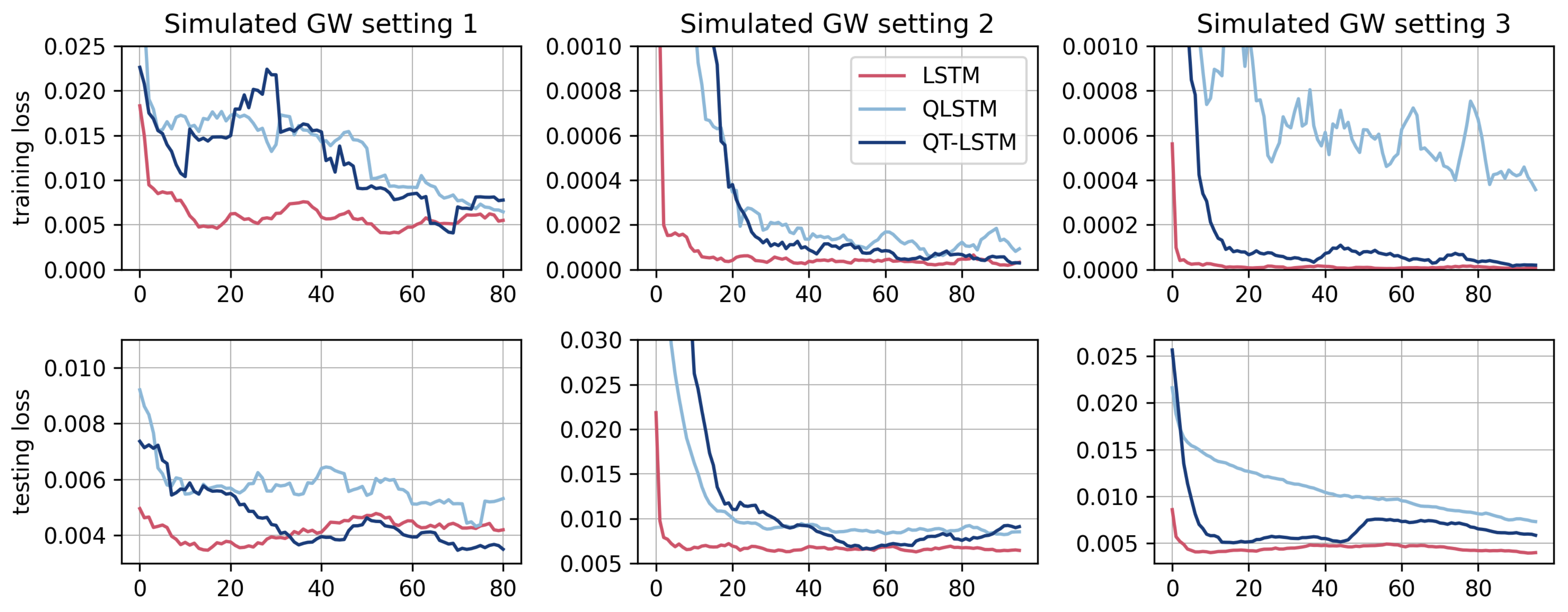}\vskip -0.1in
\caption{{\bfseries Results: Training and testing loss during federated learning scheme with different simulated GW settings for LSTM, QLSTM, and QT-LSTM. Horizontal axis represents training rounds. }}
\label{fig:simu_res_0}
\end{center}
\vskip 0.2in
\end{figure*}

\begin{figure*}[htbp]
\vskip -0.1in
\begin{center}
\includegraphics[width=1.8\columnwidth]{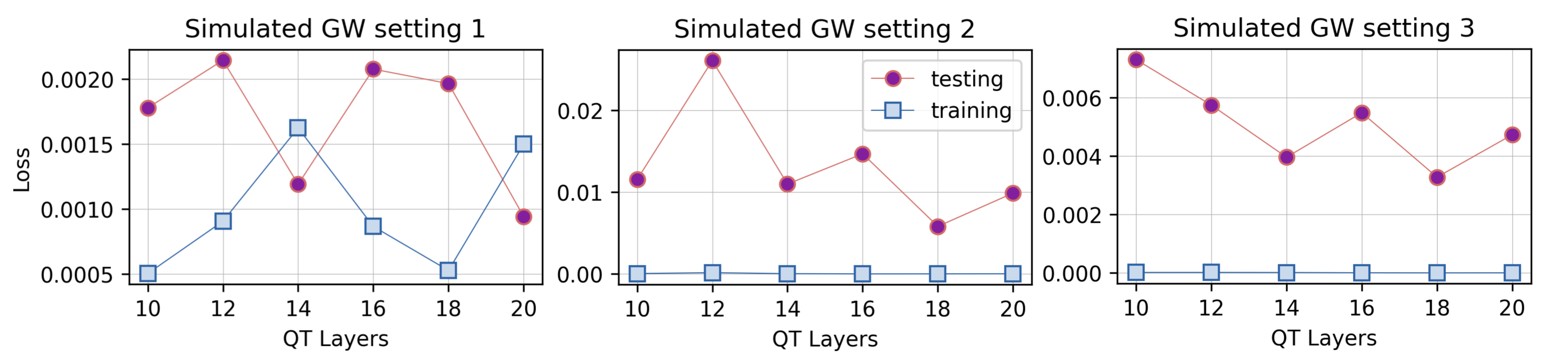}\vskip -0.1in
\caption{{\bfseries Results: Different QT layers of Fed-QT-LSTM for different simulated GW settings}}
\label{fig:simu_res_2}
\end{center}
\vskip -0.2in
\end{figure*}

\subsection{Result Analysis}

As shown in \figureautorefname{\ref{fig:simu_res_1}}, 3 different settings of GWs are shown, with prediction of Federated QT-LSTM. The main distinction of these settings is the way dividing the training data and testing data, which separated by the brown dashed line around $2/3$ of the time unit. In the left $2/3$ of time unit is the training data and the remaining $1/3$ time unit in the right is the testing data. 

\figureautorefname{\ref{fig:simu_res_0}} presents the training and testing loss observed during the federated learning process for three different simulated GW settings, comparing the performance of LSTM, QLSTM, and QT-LSTM models. The top row depicts the training loss over 100 rounds, while the bottom row shows the corresponding testing loss. Across all settings, QT-LSTM (dark blue) consistently demonstrates similar or superior performance compared to LSTM (red) and achieves lower training and testing loss than QLSTM (light blue). These results underscore the robustness and efficiency of QT-LSTM in addressing federated learning tasks for GW signal reconstruction, particularly given that the number of training parameters in QT-LSTM is only $6.9\%$ of that of LSTM, highlighting its computational efficiency without compromising performance. 

Interestingly, QLSTM exhibits the highest training and testing loss among the three models. However, it is worth noting that during the inference stage, the number of parameters in QT-LSTM is reduced to match that of the classical LSTM, resulting in 1,781 parameters for inference. In contrast, QLSTM, which utilizes a quantum computer during inference, requires only 205 parameters, significantly fewer than the other two approaches. This highlights the trade-off between leveraging quantum computing during inference (as in QLSTM) or restricting its use to the training phase (as in QT-LSTM), balancing computational efficiency and performance.

In the QT framework, the number of QT layers, denoted as $ L $, is a critical hyperparameter that determines the depth of the QNN used for model compression. Typically, deeper QNNs are more expressive and capable of capturing complex patterns.  \figureautorefname{\ref{fig:simu_res_2}} examines the impact of QT layers $ L $ on the performance across three different simulated GW settings.

For setting 1, no significant trend is observed with increasing QT layers, indicating that the task is relatively simple and can be effectively handled by a shallower QNN. In contrast, for settings 2 and 3, a clearer trend emerges, where deeper QNNs result in improved testing loss. This suggests that the tasks in these two settings are more challenging, necessitating a deeper QNN to achieve better testing performance. These findings demonstrate the importance of tailoring the QT layer depth to the complexity of the task at hand.

\section{Conclusion and Future Work}
\label{sec:cfw}


In this study, we proposed Fed-QT-LSTM, a novel integration of the QT framework with LSTM models in a FL setup. By utilizing QNNs to generate the weights of classical LSTM models, the proposed approach addresses key challenges in distributed learning systems, including model compression, scalability, and efficiency. The elimination of reliance on quantum devices during the inference stage further enhances the practicality of the framework. Our numerical simulations on GW signal datasets demonstrate that Fed-QT-LSTM consistently achieves superior performance compared to traditional LSTM and QLSTM models, achieving lower training and testing losses considering it significantly reducing the number of trainable parameters.

The study highlights the adaptability of the Fed-QT-LSTM model across different complexity levels of tasks, as observed in simulated GW settings. Specifically, the depth of the QNN (controlled by QT layers $L$) plays a critical role in balancing expressiveness and efficiency, with deeper networks required for more complex tasks.

In future work, we aim to extend the Fed-QT-LSTM framework to a broader range of datasets and application domains, including complex real-world scenarios such as weather prediction and large-scale natural simulations. 


\clearpage
\bibliographystyle{IEEEtran}
\bibliography{bib/tools,bib/vqc,bib/qml_examples,bib/quantum_fl, bib/ml_examples, bib/hybrid_co_examples,bib/classical_fl,references,bib/fwp,bib/qt}

\begin{thebibliography}{10}
\providecommand{\url}[1]{#1}
\csname url@samestyle\endcsname
\providecommand{\newblock}{\relax}
\providecommand{\bibinfo}[2]{#2}
\providecommand{\BIBentrySTDinterwordspacing}{\spaceskip=0pt\relax}
\providecommand{\BIBentryALTinterwordstretchfactor}{4}
\providecommand{\BIBentryALTinterwordspacing}{\spaceskip=\fontdimen2\font plus
\BIBentryALTinterwordstretchfactor\fontdimen3\font minus \fontdimen4\font\relax}
\providecommand{\BIBforeignlanguage}[2]{{%
\expandafter\ifx\csname l@#1\endcsname\relax
\typeout{** WARNING: IEEEtran.bst: No hyphenation pattern has been}%
\typeout{** loaded for the language `#1'. Using the pattern for}%
\typeout{** the default language instead.}%
\else
\language=\csname l@#1\endcsname
\fi
#2}}
\providecommand{\BIBdecl}{\relax}
\BIBdecl

\bibitem{biamonte2017quantum}
J.~Biamonte, P.~Wittek, N.~Pancotti, P.~Rebentrost, N.~Wiebe, and S.~Lloyd, ``Quantum machine learning,'' \emph{Nature}, vol. 549, no. 7671, pp. 195--202, 2017.

\bibitem{lau2017quantum}
H.-K. Lau, R.~Pooser, G.~Siopsis, and C.~Weedbrook, ``Quantum machine learning over infinite dimensions,'' \emph{Physical review letters}, vol. 118, no.~8, p. 080501, 2017.

\bibitem{mitarai2018quantum}
K.~Mitarai, M.~Negoro, M.~Kitagawa, and K.~Fujii, ``Quantum circuit learning,'' \emph{Physical Review A}, vol.~98, no.~3, p. 032309, 2018.

\bibitem{chen2021end}
S.~Y.-C. Chen, C.-M. Huang, C.-W. Hsing, and Y.-J. Kao, ``An end-to-end trainable hybrid classical-quantum classifier,'' \emph{Machine Learning: Science and Technology}, vol.~2, no.~4, p. 045021, 2021.

\bibitem{chen2022quantumCNN}
S.~Y.-C. Chen, T.-C. Wei, C.~Zhang, H.~Yu, and S.~Yoo, ``Quantum convolutional neural networks for high energy physics data analysis,'' \emph{Physical Review Research}, vol.~4, no.~1, p. 013231, 2022.

\bibitem{qmlapp2}
K.-C. Chen, X.~Xu, H.~Makhanov, H.-H. Chung, and C.-Y. Liu, ``Quantum-enhanced support vector machine for large-scale multi-class stellar classification,'' in \emph{International Conference on Intelligent Computing}.\hskip 1em plus 0.5em minus 0.4em\relax Springer, 2024, pp. 155--168.

\bibitem{chen2024compressedmediq}
K.-C. Chen, Y.-T. Li, T.-Y. Li, and C.-Y. Liu, ``Compressedmediq: Hybrid quantum machine learning pipeline for high-dimentional neuroimaging data,'' \emph{arXiv preprint arXiv:2409.08584}, 2024.

\bibitem{chen2020variational}
S.~Y.-C. Chen, C.-H.~H. Yang, J.~Qi, P.-Y. Chen, X.~Ma, and H.-S. Goan, ``Variational quantum circuits for deep reinforcement learning,'' \emph{IEEE access}, vol.~8, pp. 141\,007--141\,024, 2020.

\bibitem{chen2022variational}
S.~Y.-C. Chen, C.-M. Huang, C.-W. Hsing, H.-S. Goan, and Y.-J. Kao, ``Variational quantum reinforcement learning via evolutionary optimization,'' \emph{Machine Learning: Science and Technology}, vol.~3, no.~1, p. 015025, 2022.

\bibitem{yun2023quantum}
W.~J. Yun, J.~Park, and J.~Kim, ``Quantum multi-agent meta reinforcement learning,'' in \emph{Proceedings of the AAAI Conference on Artificial Intelligence}, vol.~37, no.~9, 2023, pp. 11\,087--11\,095.

\bibitem{chen2024efficient}
S.~Y.-C. Chen, ``Efficient quantum recurrent reinforcement learning via quantum reservoir computing,'' in \emph{ICASSP 2024-2024 IEEE International Conference on Acoustics, Speech and Signal Processing (ICASSP)}.\hskip 1em plus 0.5em minus 0.4em\relax IEEE, 2024, pp. 13\,186--13\,190.

\bibitem{chen2022quantumLSTM}
S.~Y.-C. Chen, S.~Yoo, and Y.-L.~L. Fang, ``Quantum long short-term memory,'' in \emph{ICASSP 2022-2022 IEEE International Conference on Acoustics, Speech and Signal Processing (ICASSP)}.\hskip 1em plus 0.5em minus 0.4em\relax IEEE, 2022, pp. 8622--8626.

\bibitem{chen2022reservoir}
S.~Y.-C. Chen, D.~Fry, A.~Deshmukh, V.~Rastunkov, and C.~Stefanski, ``Reservoir computing via quantum recurrent neural networks,'' \emph{arXiv preprint arXiv:2211.02612}, 2022.

\bibitem{lin2024quantum}
C.-H.~A. Lin, C.-Y. Liu, and K.-C. Chen, ``Quantum-train long short-term memory: Application on flood prediction problem,'' \emph{arXiv preprint arXiv:2407.08617}, 2024.

\bibitem{chen2024qcq}
K.-C. Chen, X.~Li, X.~Xu, Y.-Y. Wang, and C.-Y. Liu, ``Quantum-classical-quantum workflow in quantum-hpc middleware with gpu acceleration,'' in \emph{2024 International Conference on Quantum Communications, Networking, and Computing (QCNC)}.\hskip 1em plus 0.5em minus 0.4em\relax IEEE, 2024, pp. 304--311.

\bibitem{chen2021federated}
S.~Y.-C. Chen and S.~Yoo, ``Federated quantum machine learning,'' \emph{Entropy}, vol.~23, no.~4, p. 460, 2021.

\bibitem{chehimi2022quantum}
M.~Chehimi and W.~Saad, ``Quantum federated learning with quantum data,'' in \emph{ICASSP 2022-2022 IEEE International Conference on Acoustics, Speech and Signal Processing (ICASSP)}.\hskip 1em plus 0.5em minus 0.4em\relax IEEE, 2022, pp. 8617--8621.

\bibitem{kwak2023quantum}
Y.~Kwak, W.~J. Yun, J.~P. Kim, H.~Cho, J.~Park, M.~Choi, S.~Jung, and J.~Kim, ``Quantum distributed deep learning architectures: Models, discussions, and applications,'' \emph{ICT Express}, vol.~9, no.~3, pp. 486--491, 2023.

\bibitem{chehimi2023foundations}
M.~Chehimi, S.~Y.-C. Chen, W.~Saad, D.~Towsley, and M.~Debbah, ``Foundations of quantum federated learning over classical and quantum networks,'' \emph{IEEE Network}, 2023.

\bibitem{rofougaran2024federated}
R.~Rofougaran, S.~Yoo, H.-H. Tseng, and S.~Y.-C. Chen, ``Federated quantum machine learning with differential privacy,'' in \emph{ICASSP 2024-2024 IEEE International Conference on Acoustics, Speech and Signal Processing (ICASSP)}.\hskip 1em plus 0.5em minus 0.4em\relax IEEE, 2024, pp. 9811--9815.

\bibitem{Chehimi2024FedQLSTM}
M.~Chehimi, S.~Y.-C. Chen, W.~Saad, and S.~Yoo, ``Federated quantum long short-term memory (fedqlstm),'' \emph{Quantum Machine Intelligence}, vol.~6, no.~2, p.~43, Jul 2024.

\bibitem{chu2023cryptoqfl}
C.~Chu, L.~Jiang, and F.~Chen, ``Cryptoqfl: quantum federated learning on encrypted data,'' in \emph{2023 IEEE International Conference on Quantum Computing and Engineering (QCE)}, vol.~1.\hskip 1em plus 0.5em minus 0.4em\relax IEEE, 2023, pp. 1231--1237.

\bibitem{qu2024quantum}
Z.~Qu, L.~Zhang, and P.~Tiwari, ``Quantum fuzzy federated learning for privacy protection in intelligent information processing,'' \emph{IEEE Transactions on Fuzzy Systems}, 2024.

\bibitem{huang2021power}
H.-Y. Huang, M.~Broughton, M.~Mohseni, R.~Babbush, S.~Boixo, H.~Neven, and J.~R. McClean, ``Power of data in quantum machine learning,'' \emph{Nature communications}, vol.~12, no.~1, p. 2631, 2021.

\bibitem{liu2024training}
C.-Y. Liu, E.-J. Kuo, C.-H.~A. Lin, S.~Chen, J.~G. Young, Y.-J. Chang, and M.-H. Hsieh, ``Training classical neural networks by quantum machine learning,'' \emph{arXiv preprint arXiv:2402.16465}, 2024.

\bibitem{liu2024quantum}
C.-Y. Liu, E.-J. Kuo, C.-H.~A. Lin, J.~G. Young, Y.-J. Chang, M.-H. Hsieh, and H.-S. Goan, ``Quantum-train: Rethinking hybrid quantum-classical machine learning in the model compression perspective,'' \emph{arXiv preprint arXiv:2405.11304}, 2024.

\bibitem{liu2024qtrl}
C.-Y. Liu, C.-H.~A. Lin, C.-H.~H. Yang, K.-C. Chen, and M.-H. Hsieh, ``Qtrl: Toward practical quantum reinforcement learning via quantum-train,'' \emph{arXiv preprint arXiv:2407.06103}, 2024.

\bibitem{liu2024federated}
C.-Y. Liu and S.~Y.-C. Chen, ``Federated quantum-train with batched parameter generation,'' \emph{arXiv preprint arXiv:2409.02763}, 2024.

\bibitem{liu2024quantum2}
C.-Y. Liu, C.-H.~A. Lin, and K.-C. Chen, ``Quantum-train with tensor network mapping model and distributed circuit ansatz,'' \emph{arXiv preprint arXiv:2409.06992}, 2024.

\bibitem{lin2024quantum2}
C.-H.~A. Lin, C.-Y. Liu, S.~Y.-C. Chen, and K.-C. Chen, ``Quantum-trained convolutional neural network for deepfake audio detection,'' \emph{arXiv preprint arXiv:2410.09250}, 2024.

\bibitem{liu2024quantum3}
C.-Y. Liu, C.-H.~H. Yang, M.-H. Hsieh, and H.-S. Goan, ``A quantum circuit-based compression perspective for parameter-efficient learning,'' \emph{arXiv preprint arXiv:2410.09846}, 2024.

\bibitem{hochreiter1997long}
S.~Hochreiter, ``Long short-term memory,'' \emph{Neural Computation MIT-Press}, 1997.

\bibitem{sutskever2014sequence}
I.~Sutskever, ``Sequence to sequence learning with neural networks,'' \emph{arXiv preprint arXiv:1409.3215}, 2014.

\bibitem{li2023pqlm}
S.~S. Li, X.~Zhang, S.~Zhou, H.~Shu, R.~Liang, H.~Liu, and L.~P. Garcia, ``Pqlm-multilingual decentralized portable quantum language model,'' in \emph{ICASSP 2023-2023 IEEE International Conference on Acoustics, Speech and Signal Processing (ICASSP)}.\hskip 1em plus 0.5em minus 0.4em\relax IEEE, 2023, pp. 1--5.

\bibitem{stein2023applying}
J.~Stein, I.~Christ, N.~Kraus, M.~B. Mansky, R.~M{\"u}ller, and C.~Linnhoff-Popien, ``Applying qnlp to sentiment analysis in finance,'' in \emph{2023 IEEE International Conference on Quantum Computing and Engineering (QCE)}, vol.~2.\hskip 1em plus 0.5em minus 0.4em\relax IEEE, 2023, pp. 20--25.

\bibitem{di2022dawn}
R.~Di~Sipio, J.-H. Huang, S.~Y.-C. Chen, S.~Mangini, and M.~Worring, ``The dawn of quantum natural language processing,'' in \emph{ICASSP 2022-2022 IEEE International Conference on Acoustics, Speech and Signal Processing (ICASSP)}.\hskip 1em plus 0.5em minus 0.4em\relax IEEE, 2022, pp. 8612--8616.

\bibitem{abbas2021power}
A.~Abbas, D.~Sutter, C.~Zoufal, A.~Lucchi, A.~Figalli, and S.~Woerner, ``The power of quantum neural networks,'' \emph{Nature Computational Science}, vol.~1, no.~6, pp. 403--409, 2021.

\bibitem{yu2024shedding}
S.~Yu, Z.~Jia, A.~Zhang, E.~Mer, Z.~Li, V.~Crescimanna, K.-C. Chen, R.~B. Patel, I.~A. Walmsley, and D.~Kaszlikowski, ``Shedding light on the future: Exploring quantum neural networks through optics,'' \emph{Advanced Quantum Technologies}, p. 2400074, 2024.

\bibitem{du2020expressive}
Y.~Du, M.-H. Hsieh, T.~Liu, and D.~Tao, ``Expressive power of parametrized quantum circuits,'' \emph{Physical Review Research}, vol.~2, no.~3, p. 033125, 2020.

\bibitem{mcmahan2017communication}
B.~McMahan, E.~Moore, D.~Ramage, S.~Hampson, and B.~A. y~Arcas, ``Communication-efficient learning of deep networks from decentralized data,'' in \emph{Artificial Intelligence and Statistics}.\hskip 1em plus 0.5em minus 0.4em\relax PMLR, 2017, pp. 1273--1282.

\bibitem{cerezo2021variational}
M.~Cerezo, A.~Arrasmith, R.~Babbush, S.~C. Benjamin, S.~Endo, K.~Fujii, J.~R. McClean, K.~Mitarai, X.~Yuan, L.~Cincio \emph{et~al.}, ``Variational quantum algorithms,'' \emph{Nature Reviews Physics}, vol.~3, no.~9, pp. 625--644, 2021.

\bibitem{abbott2016observation}
B.~P. Abbott, R.~Abbott, T.~Abbott, M.~Abernathy, F.~Acernese, K.~Ackley, C.~Adams, T.~Adams, P.~Addesso, R.~X. Adhikari \emph{et~al.}, ``Observation of gravitational waves from a binary black hole merger,'' \emph{Physical review letters}, vol. 116, no.~6, p. 061102, 2016.

\bibitem{van2021riroriro}
W.~G. van Zeist, H.~F. Stevance, and J.~Eldridge, ``Riroriro: Simulating gravitational waves and evaluating their detectability in python,'' \emph{arXiv preprint arXiv:2103.06943}, 2021.

\end{thebibliography}

\end{document}